\RequirePackage{fix-cm}
\documentclass[twocolumn]{svjour3}          
\smartqed  
\usepackage{graphicx}
\usepackage{booktabs} 
\usepackage{hyperref}

\newcommand{\ba}{\mathbf{a}}
\newcommand{\bb}{\mathbf{b}}

\journalname{Eur. Phys. J. Plus}
\begin{document}

\title{Comment on "Six--body bound system calculations in the case of effective $\alpha-$core structure" [Eur. Phys. J. Plus (2016) 131: 240]}


\author{
M. R. Hadizadeh \and M. Radin \and S. Bayegan
}


\institute{
M. R. Hadizadeh \at
Institute of Nuclear and Particle Physics and Department of Physics and Astronomy, Ohio University, Athens, OH 45701, USA, \\
College of Science and Engineering, Central State University, Wilberforce, OH 45384, USA,\\
\email{hadizadm@ohio.edu} 
\and           
M. Radin \at
Department of Physics, K. N. Toosi University of Technology, P.O.Box 16315--1618, Tehran, Iran,\\
\email{radin@kntu.ac.ir}
\and   
S. Bayegan \at
Department of Physics, University of Tehran, Tehran, Iran.\\
\email{bayegan@ut.ac.ir}
}

\date{Received: date / Accepted: date}

\maketitle

\begin{abstract}

The authors argue that the calculated $^6$He binding energies by the solution of the coupled Yakubovsky integral equation in a partial wave decomposition reported by E. Ahmadi Pouya and A. A. Rajabi [Eur. Phys. J. Plus (2016) 131: 240] are incorrect. The formalism of the paper has serious mistakes and the numerical results are not reproducible and cannot be validated.

\end{abstract}

The Yakubovsky formalism for six--nucleon bound state leads to five coupled equations which can be reduced to two coupled ones for the halo structure of the two loosely bound neutrons with respect to the core nucleons \cite{Witala-FBS51}. 
Ahmadi Pouya and Rajabi have recently formulated the six--body (6B) Yakubovsky equations in a partial wave decomposition \cite{Ahmadi-EPJP131}. For simplification of the formalism, they have projected the coupled Yakubovsky components onto the $s-$wave basis states and solved the integral equations for one--term separable Yamaguchi potential to calculate the binding energy of $^6$He.

In this comment, we have shown that the formalism of the paper has serious mistakes and all the numerical results are wrong.
In the following, we have addressed few of these mistakes and flaws.

\section{Mistakes in the formalism}

\begin{enumerate}

\item How it is possible to derive the equations (20) and (22) from the equations (19) and (21), before defining a coordinate system which is discussed in section 4? The equations (20) and (22) are also inconsistent with the equations (25) and (26), as an obvious mistake, there is no azimuthal integration, i.e. $\phi_{45}$ and $\phi_{42'}$, in these equations.

\item In equations (23) and (24), the definition of the angle variables $x_{22'}$ and $X_{22'}$ are not consistent with the defined coordinate system.
As the authors have mentioned in section 4, for both Jacobi momentum vector sets, both of the second and the integration vectors are free in the space and consequently $x_{22'}$ and $X_{22'}$ should be dependent to the azimuthal angles $\phi_2$ and $\phi'_2$.

\item In equations (23) and (24), the definition of the angle variables $x_{42'}$, $x_{52'}$, $x_{45}$, $X_{42'}$, $X_{52'}$ and $X_{45}$ are incorrect. The dot products in these definition is meaningless. For example $x_{42'}$ should be defined as:
$x_{42'} \equiv \cos( \hat{\ba}_4 , \hat{\ba}_2')  = x_4 \, x_2' + \sqrt{1-x_4^2} \sqrt{1-x_2'^2} \cos(\phi'_2)$.

\item In equation (25) the shifted momentum arguments $\pi_1$ and $\pi_2$ should be exchanged in the potential form factor $g$ and also in all of the Yakubovsky components appeared in the kernel of the integral equation.

\item In equation (25), the second and third terms of the right-hand side of the integral equation, are missing a factor of $1/(4\pi)$.

\item In equation (25), the fifth and sixth terms of the right-hand side of the integral equation, are missing a factor of $1/(4\pi)^2$.

\item In equation (26), the $1/2$ factor must be applied just for the first, fourth and seventh terms of the right-hand side of the integral equation, not for all terms.

\item In equation (26), the second, third, fifth and sixth terms of the right-hand side of the integral equation, are missing a factor of $1/(4\pi)^2$.

\item In equation (A.8), the Jacobi momentum variables in the shifted momentum argument $\pi$ are exchanged. The correct shifted momentum arguments in the Kronecker delta functions are $\delta[a'_1-\pi(a'_2, a''_2)]$ and $\delta[a''_1-\pi(a''_2, a'_2)]$.
This mistake leads to a series of mistakes in the next equations, where the shifted momentum arguments $\pi_1$ and $\pi_2$ 
should be exchanged.

\item In equations (A.12) and (A.14) the factor $1/2$ should be replaced with the factor $1/(8\pi)$.

\item In equations (A.18) and (A.20) the matrix elements of the permutation operators $P_{34}P_{45}$ and $P_{34}P_{46}$ are given incorrectly. As one can see there is no azimuthal integration and clearly, they are not consistent with the selected coordinate system discussed in section 4.
Moreover, in both equations, the factor $1/2$ should be replaced with the factor $1/2 \times 1/(4\pi)^2$.

\item In equation (A.19) the definition of the shifted momentum argument $\tilde{a}_5^{*}$ should be corrected as $\tilde{a}_5^{*} = |\ba'_2 - \frac{1}{3} \ba'_3 - \ba'_4 + \frac{1}{4} \ba'_5|$.

\item In equation (A.23) the definition of the shifted momentum argument $\tilde{b}_2$ should be corrected as $\tilde{b}_2 = |\frac{1}{2}\ba'_2 - \frac{2}{3} \ba'_3 |$.

\item In equations (B.10), (B.12), (B.16) and (B.18) the matrix elements of the permutation operators $P_{45}$, $P_{46}$, $P_{34}P_{45}$ and $P_{34}P_{46}$ are given incorrectly. As one can see there is no azimuthal integration and clearly, they are not consistent with the selected coordinate system discussed in section 4. Moreover, the factor $1/2$ should be replaced with the factor $1/2 \times 1/(4\pi)^2$.

\item In equation (B.19) the definition of the shifted momentum argument $\tilde{b}_2^{**}$ should be corrected as $\tilde{b}_2^{**} = \frac{2}{3}|\bb'_2 + \bb'_3 |$.

\item In equation (B.20) the Kronecker delta functions in the kernel of the integral are incorrect. They should be as $\delta[a'_2-\bar{a}_2]$ and $\delta[a'_3-\bar{a}_3]$.

\end{enumerate}

In summary, the published formalism has serious mistakes which can completely change the numerical results of the solution of the coupled Yakubovsky integral equations. The mistakes in the formalism can be easily verified by simplification of the problem to a four-- or three--body bound state.
Clearly, the published formalism cannot reproduce the partial wave representation of Faddeev and Yakubovsky equations given in Refs. \cite{pw-0} and \cite{pw-1}.

\section{Mistakes in the numerical results}

The authors have used one--term separable Yamaguchi potential to solve the coupled Yakubovsky integral equations and they have reported a 6B binding energy of $-92.34$ MeV. 
The authors have not discussed in their paper about the numerical issues and challenges, like:
\begin{itemize}
\item The mapping they have used for the magnitude of the Jacobi momentum vectors,
\item the momentum cutoffs they have used in the solution of the integral equations,
\item the number of iterations for the solution of the coupled integral equations,
\item the runtime and parallelization algorithm for the solution of the coupled integral equations,
\item the plots of the Yakubovsly components to verify the halo structure of $^6He$.
\end{itemize}

To verify the reported 6B binding energy, we have solved the coupled Yakubovsky integral equations, of course by considering the mentioned corrections in the formalism.
In our calculations we have used a hyperbolic--linear mapping for the magnitude of the Jacobi momentum vectors on the domain $[0,5] \cup [5,10] \cup [10,50]$ fm$^{-1}$ and for the construction of the orthonormal basis in Lanczos technique we have used
seven iterations.

As we have shown in Table \ref{6B}, the solution of the coupled Yakubovsky integral equations for 6B bound state using $N_{jac}=20$, $N_{sph}=14$ and $N_{pol}=14$ doesn't even converge, whereas the authors of Ref. \cite{Ahmadi-EPJP131} have reported the 6B binding energy of $-92.34$ MeV.

\begin{table}[hbt] 
\begin{center}
\begin{tabular}{cccccc|ccccccccc}
\toprule
$N_{jac}$& $N_{sph}$  & $N_{pol}$&  $E_6$ (MeV)& $E_6$ (MeV) Ref. \cite{Ahmadi-EPJP131} \\
\hline
10 & 6 & 6& -118.23550   & --  \\
10 & 10 & 10& -115.75205  & --     \\
10 & 14 & 14& -115.76635  & -- \\
14 & 14 & 14& No Convergence & --   \\
20 & 14 & 14& No Convergence  & -92.34  \\
\bottomrule
\end{tabular}
\end{center}
\caption{The convergence of the six--body binding energy for Yamaguchi I potential as a function of the number of the grid points. $N_{jac}$ is the number of mesh points for the magnitude of the Jacobi momentum vectors, $N_{sph}$ is the number of mesh points for the spherical angles and $N_{pol}$ is the number of mesh points for the azimuthal angles.}
\label{6B}
\end{table}%

As a second test, we have verified the stability of the eigenvalue $\eta$ as a function of the number of the grid points.
As we have shown in Table \ref{eta}, the largest positive eigenvalue obtained from the solution of the coupled Yakubovsky integral equations is so sensitive to the number of grid points for the magnitude of the Jacobi momentum vectors, i.e. $N_{jac}$, and of course are quite different from the published eigenvalues in Ref. \cite{Ahmadi-EPJP131}.

\begin{table}[hbt] 
\begin{center}
\begin{tabular}{cccccccccccc}
\toprule
$N_{jac}$& $N_{sph}$  & $N_{pol}$&  $\eta$ &  $\eta$ Ref. \cite{Ahmadi-EPJP131} \\
\hline
10 & 6 & 6& 1.03945  & -- \\
10 & 10 & 10& 1.03541  & --\\
10 & 14 & 14& 1.03545&  0.973 \\
14 & 14 & 14&  0.67553  & -- \\
20 & 14 & 14& 1.28076  &  1.000  \\
\bottomrule
\end{tabular}
\end{center}
\caption{The largest positive eigenvalue for the input six--body binding energy of $-92.34$ MeV for Yamaguchi I potential as a function of the number of the grid points. $N_{jac}$ is the number of mesh points for the magnitude of the Jacobi momentum vectors, $N_{sph}$ is the number of mesh points for the spherical angles and $N_{pol}$ is the number of mesh points for the azimuthal angles.}
\label{eta}
\end{table}%

We believe the above--mentioned mistakes are quite enough to ensure us that the authors have reported not genuine results.
Similar to other poor paper published by Ahmadi Pouya and Rajabi \cite{Ahmadi-IJMPE25} about the solution of the six--body Yakubovsky equations in a three--dimensional approach, as we have discussed in another comment \cite{Hadizadeh-comment}, not only the formalism of the paper has serious mistakes, but also the numerical results are not trustable.


\end{document}